# Numerical simulation of leakage effect for quantum NOT operation on three-Josephson-junction flux qubit

[1]Tao Wu，[1]Jianshe Liu, [2]Zheng Li

[1]Institute of Microelectronics, Tsinghua University, Beijing 100084

[2]Department of Electronic Engineering, Tsinghua University, Beijing 100084

Superconducting flux qubits with three Josephson junctions are promising candidates for the building blocks of a quantum computer. We have applied the imaginary time evolution method to study the model of this qubit accurately by calculating its wave functions and eigenenergies. Because such qubits are manipulated with magnetic flux microwave pulses they might be irradiated into non-computational states which is called the leakage effect. Through the evolution of the density matrix of the qubit under either hard-shaped π-pulse or Gaussian-shaped π-pulse to carry out quantum NOT operation，it has been demonstrated that the leakage effect for a flux qubit is very small even for hard-shaped microwave pulses while Gaussian-shaped pulses may suppress the leakage effect to a negligible level.

Superconducting qubits[1, 2] are solid-state macroscopic systems conforming to the principles of quantum mechanics at ultra low temperatures. They can be compatibly fabricated in a microelectronic process line and easily manipulated by on-chip microwave currents or flux pulses, which make them promising candidates for the building blocks of a quantum computer.[3] For superconducting flux qubits,[4-9] the magnetic flux is the convenient parameter to mark and control their eigenstates. Flux qubits are insensitive to background charge fluctuations but relatively fragile to magnetic flux noise compared to superconducting charge[10, 11] and phase[12, 13] qubits. Quantum superposition in the spectroscopy[6] and Rabi oscillations of a flux qubit in the time domain[7] have been observed.

Leakage effect[13-15] of a flux qubit during quantum operations means that the qubit escapes into non-computational subspaces. This effect of phase qubits has been discussed,[13, 14] and inspired by Lin in Ref. [15] we numerically study it for a flux qubit in this letter.

This work is based upon the calculation of the eigenstates of a single flux qubit using the imaginary time evolution method[16-19] and the evolution of its density matrix under magnetic flux pulse perturbation. It has been revealed that the populations irradiated to the second and third excited states are extremely small for Gaussian-shaped pulses and also unimportant for hard-shaped pulses, thus there is no need to consider other excited states because the leakage to higher energy levels are much smaller and therefore negligible.

A three-Josephson-junction (3JJ) flux qubit is composed of a superconducting loop interrupted by three Josephson junctions[4-6] as shown in Fig.1s. Junctions 1 and 2 have equal areas while junction 3 is $\alpha$ ($0<\alpha<1$) times smaller. The critical currents for them are $I_c$, $I_c$ and $\alpha I_c$, respectively. The loop is biased by a magnetic flux $f\Phi_0$, where $f$ is the flux frustration, $\Phi_0=h/(2e)$ is the superconducting flux quantum, $e$ is the electron charge, and $h$ is Planck constant. The qubit can be reduced to a two-level system when $f$ is in the vicinity of $n+1/2$ with $n$ an integer. Neglecting the loop inductance, the Hamiltonian of the system can be written as[5]

$$H = \frac{1}{2}\vec{p}^T \bullet M^{-1} \bullet \vec{p} + E_J[2+\alpha - \cos(\varphi_1) - \cos(\varphi_2) - \alpha\cos(2\pi f + \varphi_1 - \varphi_2)], \quad (1)$$

where $M = C_J\left(\frac{\Phi_0}{2\pi}\right)^2 \begin{pmatrix} 1+\alpha & -\alpha \\ -\alpha & 1+\alpha \end{pmatrix}$, $\vec{p} = M\vec{\dot{\varphi}}$, i.e., $(p_1 \quad p_2)^T = M(\dot{\varphi}_1 \quad \dot{\varphi}_2)^T$ and $E_J = I_c\Phi_0/(2\pi)$ is the Josephson energy for junction 1 and 2. Eq. (1) can be expressed explicitly as

$$H = \frac{(1+\alpha)}{2(1+2\alpha)C_J}\left(\frac{2\pi}{\Phi_0}\right)^2 (p_1^2 + p_2^2 + \frac{2\alpha p_1 p_2}{1+\alpha})$$
$$+ E_J[2+\alpha - \cos(\varphi_1) - \cos(\varphi_2) - \alpha\cos(2\pi f + \varphi_1 - \varphi_2)]. \quad (2)$$

It is different from the usual one as in Ref. 5 because the Hamiltonian remains in the original frame without transformation.

We have utilized the fourth order imaginary time evolution method[16-19] to calculate the above Hamiltonian. Table I shows the eigen-energies of the qubit for $f$ =0.50 and $f$= 0.495. Figs. 2A-B illustrate the ground state $|0\rangle$ and first excited state $|1\rangle$ for $f$=0.50, Figs. 2C-D show $|0\rangle$ and $|1\rangle$ for $f$=0.495, and Figs. 2E-F show the second excited state $|2\rangle$ and third excited state $|3\rangle$ for $f$=0.495. We choose in the computation $\alpha$=0.80, $E_J$=198.9437 GHz and $C_J$=7.765 fF.

**Table I.** The energies $E_i$ for the lowest four eigenstates with two values of $f$. The energies are in units of GHz.

| $f$ | $E_0$ | $E_1$ | $E_2$ | $E_3$ |
| --- | --- | --- | --- | --- |
| 0.50 | 0 | 0.3313 | 28.9984 | 35.6331 |
| 0.495 | 0 | 8.7854 | 31.9617 | 40.9154 |

The wave functions of $|0\rangle$ and $|1\rangle$ for $f$=0.50 are symmetric and anti-symmetric, respectively, and they quickly lose the symmetry when $f$ deviates from this degenerate point. To achieve larger supercurrents for readout, we bias the qubit at $f$=0.495, where the wave functions of $|0\rangle$ and $|1\rangle$ appear localized in two separate wells as shown in Figs. 2C and 2D.

Based on these results, we have studied the leakage effect during a quantum NOT operation upon one flux qubit with microwave magnetic flux pulse operations. The interaction term $W(t)$ between the microwave field and the qubit can be considered as a perturbation on the original Hamiltonian in Eq. (2), or an external microwave magnetic flux perturbation $f_\mu(t)\cos(\omega_0 t)\Phi_0$ on the flux bias $f\Phi_0$ threading the loop, i.e.,

$$W(t) \cong 2\pi\alpha E_J f_\mu(t)\cos(\omega_0 t)\sin(2\pi f + \varphi_1 - \varphi_2), \quad (3)$$

where $|f_\mu(t)| \ll f$ and $\omega_0 = (E_1 - E_0)/\hbar$. From now on, we denote by $\hbar\omega_0$ the energy unit. In the Hilbert space spaned by the eigenstates $|0\rangle$, $|1\rangle$, $|2\rangle$ and $|3\rangle$, the interaction $W(t)$ can be written as a $4\times 4$ matrix: $W = F(\tau)M\cos(\tau)$, where $\tau = \omega_0 t/(2\pi)$, $F(\tau) = -2\pi\alpha E_J f_\mu(\tau)$, and $M$ is a 4×4 matrix with $M_{i,j} = \langle i-1|\sin(2\pi f + \varphi_1 - \varphi_2)|j-1\rangle$ for $i,j$=1,2,3,4. Note that $F(\tau)$ is controllable in experiments. In the computational space, the population inversion between the ground state and the first excited state depends on the effective matrix element $M_{1,2}$ and $M_{2,1}$, i.e., 0.0339. A quantum NOT operation can be achieved by a microwave π-pulse with duration $\tau_p$, which requires

$$\int_0^{\tau_p} F(\tau)M_{1,2} d\tau = 1/2. \quad (4)$$

The hard-shaped pulse has constant microwave amplitude, and Eq. (4) yields $F(\tau) = 1/(2\tau_p M_{1,2})$ for $0 \leq \tau \leq \tau_p$. For the Gaussian-shaped pulse,[15,16] one has $F(\tau) = (1/2M_{1,2})(1/\sqrt{2\pi}\tau_w)\exp(-(\tau-\tau_p/2)^2/2\tau_w)$ for $0 \leq \tau \leq \tau_p$, where $\tau_w$ is the characteristic width of the Gaussian pulse chosen between $0.167\tau_p$ and $0.100\tau_p$.[20]

The evolution of the qubit under irradiation can be described in the interaction picture by the Liouville equation:[21]

$$i\partial\rho(\tau)/\partial\tau = 2\pi[H_1(\tau), \rho(\tau)]. \quad (5)$$

Here $H_1(\tau)$ is a matrix with $H_1(\tau)_{i,j} = \exp(i2\pi(k_i - k_j)\tau)W_{i,j}$, where $k_i = (E_{i-1} - E_0)/(\hbar\omega_0)$. Choosing a time step small enough (e.g., $\Delta\tau \approx 10^{-3}$), we can integrate the equation to obtain the density matrix for a given time. The element $\rho_{i,i}$ at the end denotes the probability $P_i$ of the system in state $|i\rangle$, $i$=1, 2, 3, 4. Table II shows the final values of the leaking populations for three time durations of pulses. Fig. 3 illustrates the leakage effect during the operations for $\tau_p$=400 (i.e., 45.5*ns*).

**Table II.** The populations ($P_i$) of the lowest four levels after the microwave magnetic flux irradiations with Hard-shaped pulses (on the left of "/") and with Gaussian-shaped pulses (on the right of "/").

|       | $\tau_p$=100 | $\tau_p$=200 | $\tau_p$=400 |
|-------|--------------|--------------|--------------|
| $P_2$ | 1.3e-5/4.6e-8 | 2.5e-6/1.4e-8 | 1.5e-6/4.2e-9 |
| $P_3$ | 2.9e-6/5.0e-8 | 1.0e-6/1.0e-8 | 3.3e-7/1.8e-9 |

It is revealed that Gaussian-shaped π-pulse inhibit leakage effects 100 better than hard-shaped π-pulse, which is as remarkable as that in the case of phase qubits.[13] Also, longer durations reduce the leakage. However, leakage is very small during this operation even for hard-shaped pulses.

In conclusion, we have solved the eigen-functions and eigen-energies of a 3JJ flux qubit by the imaginary time evolution method and studied the leakage effect during the quantum NOT operation through the evolution of its density matrix. It has been demonstrated that the leakage effect for a flux qubit is very small even for hard-shaped microwave pulses while Gaussian-shaped pulses may suppress the leakage effect to a negligible level.

We gratefully appreciate Johnson P R and Strauch F W in NIST for their instructions about the imaginary time evolution method. Discussions with Wang Ji-lin, Li Tie-fu, Chen Pei-yi, Yu Zhi-ping and Li Zhi-jian are acknowledged. This work is supported by the 211 Program of Nanoelectronics of Tsinghua University (210605001).

with $\tau_p = k\tau_w$ and $k \approx 6\Box 10$.
[21] Blum Karl 1981 *Density Matrix Theory and Applications* (NewYork: Plenum Press)

**Figure Captions**

**Fig. 1.** One flux qubit with crosses representing Josephson junctions. $C_{J1}$, $C_{J2}$ and $C_{J3}$ are the equivalent capacitances of the junctions with $C_{J1}=C_{J2}=C_J$ and $C_{J3}=\alpha C_J$.

**Fig. 2.** (A-B)Ground state $|0\rangle$ and first excited state $|1\rangle$ for $f$=0.50. (C-F)$|0\rangle$, $|1\rangle$, second excited state $|2\rangle$ and third excited state $|3\rangle$ for $f$=0.495.

**Fig. 3** Population of $|0\rangle$ ($P_0$, solid line), $|1\rangle$ ($P_1$,dashed line), $|2\rangle$ ($P_2$, solid line) and $|3\rangle$ ($P_3$, dashed line) after (A-B) the hard-shaped π-pulse and (C-D) the Gaussian-shaped π-pulse. The pulse duration is $\tau_p$=400.

**Fig. 1**

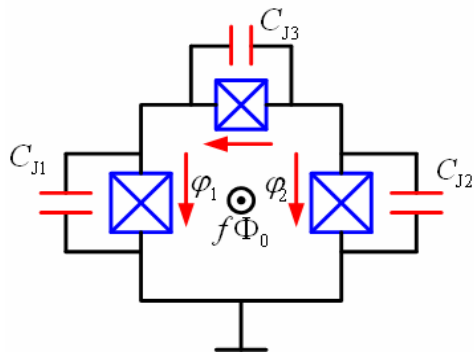

**Fig. 2**

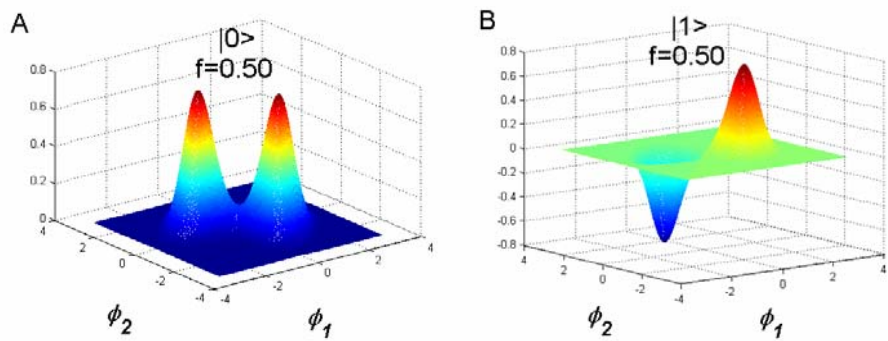

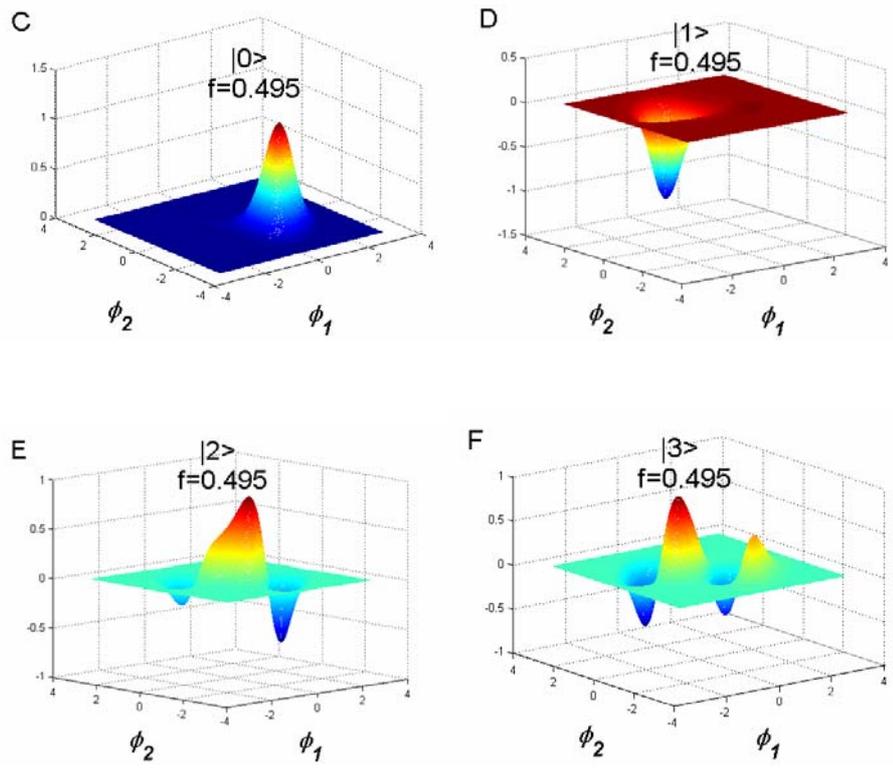

**Fig. 3**

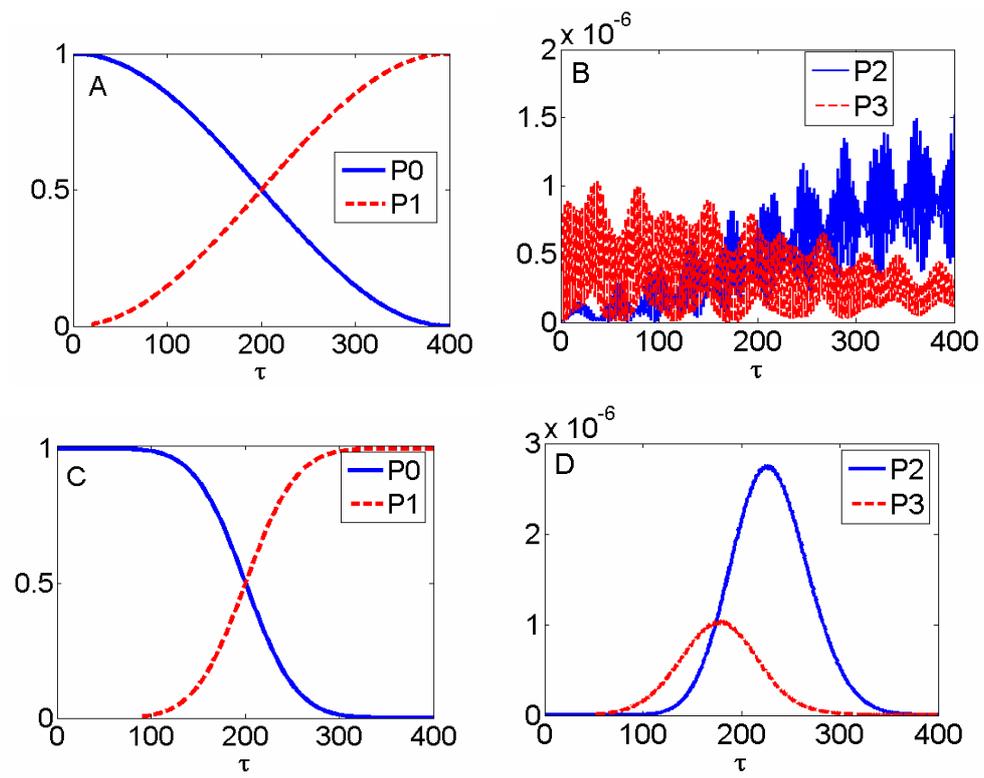